\documentstyle[aps,amssymb,amsfonts,psfig,preprint]{revtex}
\begin{document}
\tightenlines
\title{Improving quantum interferometry by using entanglement} 
\author{G. M. D'Ariano$^{a,b,c}$, Matteo G. A Paris$^a$, and Paolo Perinotti$^d$} 
\address{$^a$ Quantum Optics \& Information Group,
Istituto Nazionale di Fisica della Materia, Unit\`a di Pavia,
Dipartimento di Fisica ``A. Volta'', via Bassi 6, I-27100 Pavia, Italy\\
$^b$ Istituto Nazionale di Fisica Nucleare, Sezione di Pavia \\
$^c$ Department of Electrical and Computer Engineering, Northwestern
University, Evanston, IL  60208\\
$^d$ Dipartimento di Fisica and Sezione INFN, Universit\'a di Milano
via Celoria 16, I-20133 Milano, ITALIA}
\date{\today}\maketitle
\begin{abstract} We address the use of entanglement to improve the precision 
of generalized quantum interferometry, {\em i.e.} of binary measurements 
aimed to determine whether or not a perturbation has been applied by a 
given device. For the most relevant operations in quantum optics, we evaluate
the optimal detection strategy and the ultimate bounds to the minimum detectable
perturbation. Our results indicate that entanglement-assisted strategies
improve the discrimination in comparison with conventional schemes. 
A concrete setup to approach performances of the 
optimal strategies is also suggested. \end{abstract}
\section{Introduction}\label{s:intro}
Any interferometric setup is devised to reveal minute perturbations to a
given configuration. Such perturbations may be induced by the environment 
or by the action of a given device. In an interferometer, the internal quantum 
operation is monitored by probing the output state, which, in turn, 
results from the evolution of a given input. By suitably choosing the input 
signal and the detection stage one optimizes the interferometric measurement. 
Optimization has two main goals: i) to maximize the probability of revealing 
a perturbation, when it occurs, and ii) to minimize the value of the smallest 
perturbation that can be effectively detected.
\par
\begin{figure}[h]
\psfig{file=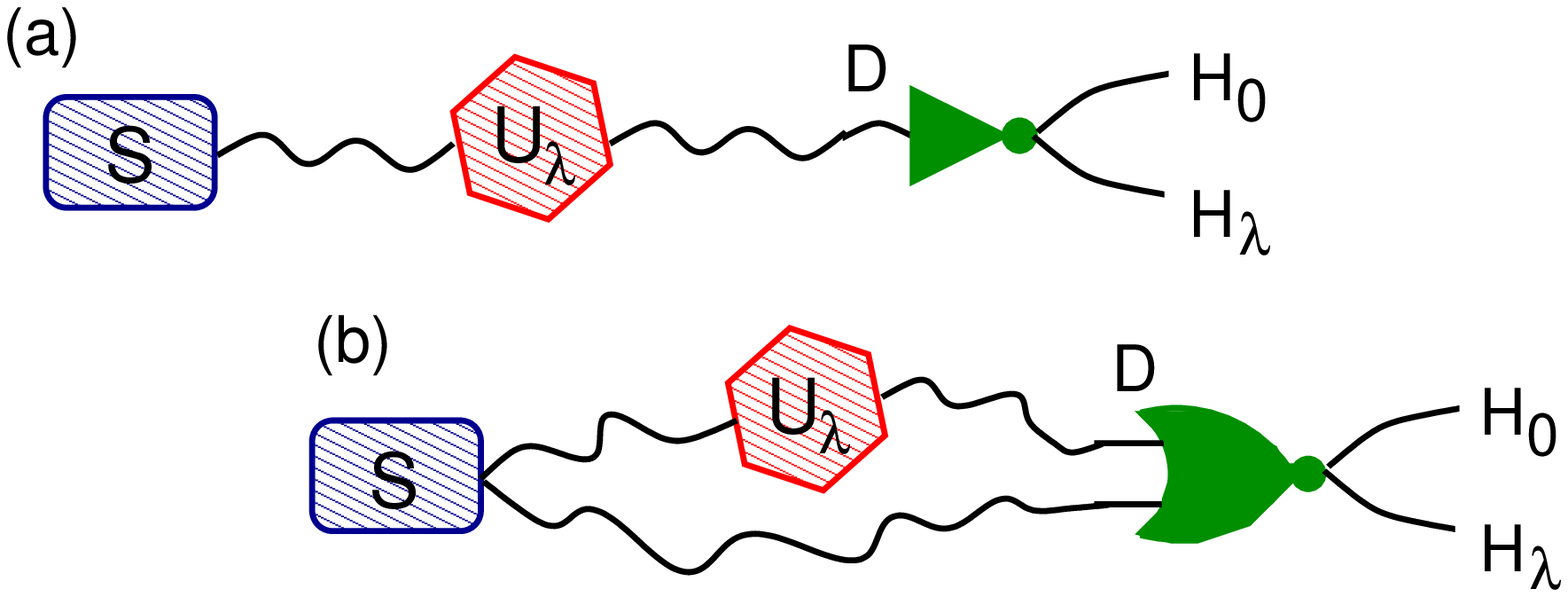,width=14cm}
\caption{A generalized interferometer is a binary detection scheme aimed 
to check whether or not a given quantum device (the hexagon in the figure) 
has performed the quantum operations described by the unitary operator 
$U_\lambda$. The signal employed as a probe is prepared by the source {\sf S} 
and then enters the device, which may or may not apply $U_\lambda$. The two 
hypothesis: ${\cal H}_0$ (the signal is unperturbed) and ${\cal H}_\lambda$ ($U_\lambda$ 
has been applied) should be discriminated on the basis of the outcome of the 
detector {\sf D}. (a): simple scheme involving a single-mode probe.  (b): 
scheme involving entanglement-assisted binary detection.}\label{f:schemes}
\end{figure}

In essence, any interferometric scheme may be viewed as a binary
communication system \cite{hollenhorst,intb}, with the perturbation playing 
the role of the encoded information. In order to see better this analogy 
let us consider the scheme shown in Fig. \ref{f:schemes}a. A source 
{\sf S} of quantum state prepares the input signal, say $\varrho_0$, 
which travels along the interferometer, and it is eventually measured by
some detector, denoted by {\sf D}. The detector is described by an operator-valued 
probability measure (POVM) $\Pi(x)$, with $x \in X$, $X$ being the manifold describing 
the possible detection outcomes. Inside the interferometer we a have generic 
quantum device, which may or may not perturb the signal, {\em i.e.} it
performs the quantum operation described by the unitary $U_\lambda$. If a perturbation 
occurs the signal is modified and, at the output, we have the state 
$\varrho_\lambda = U_\lambda\:\varrho_0 U^\dag_\lambda$. The aim of
the detection stage is to discriminate between $\varrho_0$ and its perturbed version 
$\varrho_\lambda$. An optimized interferometer is a device that is
able to tell which $\varrho$, for $\lambda$ as small as possible. 
Posed in this way, interferometry is naturally viewed as a binary decision problem, 
and the detection stage can be described by a two-value POVM $\{\Pi_0,\Pi_\lambda 
\equiv{\mathbb I}-\Pi_0\}$, which corresponds to the two possible inferences. 
\par
The main goal of the present paper is to demonstrate the benefits of
entanglement in binary interferometry. We will show that
distinguishability of the two hypothesis (${\cal H}_0$: nothing happened and 
${\cal H}_\lambda$: a perturbation has occurred) can be improved by: i) 
using an input signal which is entangled with another subsystem, 
and ii) measuring the two systems jointly at the output of the
interferometer (see Fig. \ref{f:schemes}b). 
\par
In order to optimize the detection strategies, and to show the benefits of
entanglement, we will make use of results and methods from quantum detection
theory applied to binary decision \cite{hel,hol}. This approach is particularly
useful for our purposes, since it does not refer to any specific detection
scheme for the final stage of the interferometer, but rather, owing to
its generality, it allows to find the ultimate quantum limits to
interferometry for specific classes of quantum signals.
\par
In the next Section, in order to establish notation, we briefly review 
the Neyman-Pearson approach to quantum binary decision, and state a lemma about 
minimum input-output overlap. Then, in Section 
\ref{s:core} we apply these results to the interferometric detection of
perturbations induced by the most relevant operations in quantum optics such
as displacement, squeezing, mixing and phase-shifting. As we
will see, entanglement-assisted interferometers provide better discrimination 
than conventional schemes. In Section \ref{s:diff} we analyze an 
interferometric configuration that achieves, for the
quantum operations discussed in Section \ref{s:core}, the ultimate bounds to 
precision. Finally, in Section \ref{s:outro} we close the paper with 
some concluding remarks
\section{Quantum binary decisions in the Neyman-Pearson approach} 
\label{s:qbd}
The problem that we are facing is to decide among two hypotheses
${\cal H}_0$ and ${\cal H}_1$ about the state of a system, which is
described by a density operator $\varrho$ on the Hilbert space of the
system. To each hypotheses it will correspond a different density
operator as follows 
\begin{eqnarray}
\begin{array}{cc}
{\cal H}_0:       & \hbox{the system is in the state } \varrho_0 ,\\
{\cal H}_\lambda: & \hbox{the system is in the state } \varrho_\lambda .
\end{array}
\label{hyps}
\end{eqnarray}
Of course, there are many different measurements which can provide information 
about the state of the system: each of them, however, can be recast mathematically 
as a two-value POVM, corresponding to the two possible inferences 
${\cal H}_0$ and ${\cal H}_1$, namely
\begin{equation}
\Pi_0 ,\Pi_\lambda \geq 0\qquad \Pi_0 +\Pi_\lambda ={\mathbb I}.
\end{equation}
One then needs an optimization strategy in order to determine the most reliable
measurement discriminating between the two states.
If $\varrho_0$ and $\varrho_\lambda$ are orthogonal i.e. $\varrho_0
\varrho_\lambda=\varrho_\lambda \varrho_0=0$ the solution is trivial, since $\Pi_0$ 
is the projection into any subspace that contains the support of
$\varrho_0$ and is orthogonal to the support of $\varrho_\lambda$, and
$\Pi_\lambda$ is simply the complement $\Pi_\lambda={\mathbb I}
-\Pi_0$. In most cases of interest, however, the states are not
orthogonal and one has to apply an optimization strategy.  Since
interferometric schemes are frequently used for detecting low-rate
events, we may want to look for a strategy that keeps a low-rate of
false alarm, namely of wrong inference of perturbation occurrence.
For this purpose, it is suitable to adopt a so-called Neyman-Pearson
(NP) detection strategy, which consists in fixing a tolerable 
value of the false-alarm  probability $Q_0$---the probability of inferring
that the state of the system is $\varrho_\lambda$ while it is actually
$\varrho_0$---and then maximizing the detection probability $Q_\lambda$, i.e.
the probability of a correct inference of hypothesis ${\cal H}_\lambda$ \cite{NP}. It
has been proved by Helstrom \cite{hel} and Holevo \cite{hol} that this problem
can be solved by diagonalizing the operator
\begin{equation}
\varrho_\lambda -\mu\varrho_0\:, \label{lambda}
\end{equation}
$\mu$ playing the role of a Lagrange multiplier accounting for the
bound of fixed false alarm probability. According to \cite{hel} the optimal POVM is
the one in which $\Pi_\lambda$ is the projection onto the eigenspaces of
(\ref{lambda}) relative to positive eigenvalues and $\Pi_0 ={\mathbb
I}-\Pi_\lambda$. Unfortunately, the diagonalization of (\ref{lambda})
is generally not easy. However, when $\varrho_0=|\psi_0\rangle\langle\psi_0|$ 
and $\varrho_\lambda=|\psi_\lambda\rangle\langle\psi_\lambda|$ are pure states it 
can be easily solved analytically, by expanding $|\psi_0\rangle$ and
$|\psi_\lambda \rangle$ on the eigenvectors of the difference operator
(\ref{lambda}). In this way one can evaluate both $Q_0$ and
$Q_\lambda$  versus $\mu$, and after eliminating $\mu$ from their 
expressions one obtains
\begin{equation}
Q_\lambda=\cases{\left[\sqrt{Q_0 |\kappa|^2}+\sqrt{(1-Q_0)(1-|\kappa|^2)}\right]^2 & 
	  \text{for $0\leq Q_0\leq |\kappa|^2$,}\cr 
1 & \text{for $|\kappa|^2 < Q_0\leq1$.}\cr}\label{detection}
\end{equation}
where $|\kappa|^2=|\langle\psi_0 |\psi_\lambda\rangle|^2=|\langle\psi_0 
|U_\lambda |\psi_0\rangle|^2$  is the overlap 
between the two states. The detection probability is a decreasing function 
of the overlap---the smaller the overlap, the easier the
discrimination---since one can reach detection probability $1$ while keeping
a low false alarm probability. On the contrary, when the overlap approaches 
$1$ one is forced to decrease the detection probability in order to 
keep the false alarm probability small.
\par
The optimal choice of the probe that minimizes the overlap, depends on
the eigenvalues of the unitary operation $U_\lambda$. In order to illustrate
this, let us expand $U_\lambda$ in terms of its eigenvectors $U_\lambda=
\sum_j \: e^{i \varphi_j} \: |\varphi_j \rangle\langle \varphi_j|$ (with
integrals replacing sums in case of continuous spectrum) and let's
denote by ${\cal
O}(U_\lambda)=\min_{\psi} |\langle\psi|U_\lambda |\psi\rangle|^2$ the minimum
overlap between the two possible outputs, as obtained by varying the probe
state. Then we have the following {\em overlap Lemma} \cite{part,entame}:
the minimum overlap ${\cal O}(U_\lambda)$ is given by the distance 
from the origin in the complex plane of the polygon whose vertices are
the eigenvalues of $U_\lambda$. Therefore, the overlap is either zero
(if the polygon includes the origin) or it is given by 
\begin{eqnarray}
{\cal O}(U_\lambda) = \cos^2 \frac{\Delta\varphi}{2}
\label{minO}\;,
\end{eqnarray}
where $\Delta\varphi$ is the angular spread of the eigenvalues.
Zero overlap can be achieved with a probe state that is given by a superposition 
of at least three eigenvectors of $U_\lambda$, corresponding to
eigenvalues that make a polygon that encloses the origin (or, if they exist, 
by a superposition of two of them corresponding to diametrically opposed eigenvalues). 
Instead, if the minimum overlap is not zero, it is achieved by the optimal probe
state given by 
\begin{eqnarray}
|\psi\rangle= \frac{1}{\sqrt{2}} \left(|\varphi_{i}\rangle + |
\varphi_{j}\rangle \right)
\label{minOstate}\;,
\end{eqnarray} with $\Delta\varphi=\varphi_i-\varphi_j$.
\section{Entanglement in quantum interferometry}\label{s:core}
In this section we compare the performances of single-mode (Fig. \ref{f:schemes}a) 
and entanglement-assisted interferometric schemes (Fig. \ref{f:schemes}b) in the 
detection of small perturbations induced by relevant quantum optical operations 
such as displacement, squeezing, mixing and phase-shifting. 
The comparison is made in terms of the detection sensitivity, namely,
upon parametrizing the ``size'' of the perturbation---whence the 
corresponding output state---by a coupling parameter $\lambda$. In other words, 
the comparison is made in terms of the minimum detectable value
$\lambda_{\min}$ of $\lambda$ corresponding to output states that can
be effectively discriminated while keeping the {\em acceptance ratio}
$\gamma^\star$ of the NP strategy large, namely $\gamma^\star\doteq
Q_{\lambda}/Q_0 \gg 1$. We will call the quantity $\lambda_{min}$
the ``sensitivity'' of the interferometric scheme. Using
Eq. (\ref{detection}) the above condition can be written in term of
the overlap as follows  
\begin{eqnarray}
|\kappa|^2&=&1 - \Lambda(Q_0,\gamma^\star)\label{KLmin}  \\
\Lambda(Q_0,\gamma^\star)&=&
Q_0 \left[ 1+ \gamma^\star (1-2Q_0) - 2 \sqrt{\gamma^\star 
(1-Q_0)(1-\gamma^\star Q_0)}\right]
\nonumber\;.
\end{eqnarray}
For each class of transformations, we will make some general considerations
and then focus our attention on sensitivity bounds that can be achieved using
realistic ({\em i.e.} feasible with current technology) probe signals.
\subsection{Perturbation made of a single-mode complex displacement}
Let us first consider the case when the perturbation is imposed by the
displacement operator $U_\alpha\equiv D(\alpha)=\exp (\alpha a^\dag - \bar\alpha a)$.  
In principle, in this case, the discrimination can be done exactly with 
single-mode probe. This can be seen by writing the displacement as $U_\alpha 
=\exp (i 2 |\alpha| x_\theta)$, $ x_\theta=1/2 (a^\dag e^{i \theta} + 
a e^{-\i\theta})$ being the quadrature
operator, and $\theta=\arg(\alpha)$.  Since the spectrum of the
quadrature coincides with the real axis, the spectrum of $U_\alpha$
covers the whole unit circle, and, therefore, the states $|\psi_0\rangle$ and
$|\psi_\alpha\rangle = U_\alpha |\psi_0\rangle$ can be discriminated
with certainty either by choosing $|\psi_0\rangle$ as the eigenstate of the
conjugated quadrature $x_{\theta+\pi/2}$, or, according to the overlap lemma,
as a superposition of at least two eigenstates of the quadrature $x_\theta$.
Unfortunately, such optimal states are unphysical, since they are not
normalizable and have infinite energy. Moreover, even though we
approximate them with physical states with finite energy, the
identification of the optimal states would require the knowledge of
the phase of the perturbation. In order to see that, let us rewrite the
eigenvector $|0\rangle_{\theta+\pi/2}$ as the limiting case of a
squeezed vacuum, $|0\rangle_{\theta+\pi/2}=\lim_{|\zeta| \rightarrow \infty}
|\zeta\rangle =\lim_{|\zeta| \rightarrow \infty} S(\zeta) |0\rangle $,
where $\theta=\arg(\zeta)+\pi/2$ is the argument of the squeezing
parameter $\zeta$ of the squeezing operator given by
\begin{equation}
S(\zeta) =\exp [1/2(\zeta^2 a^{+2} - \bar\zeta a^2)]\label{sqU},
\end{equation}
and $|0\rangle$ is the electromagnetic vacuum. Our squeezed vacuum has
mean photon number $N=\sinh^2 |\zeta|$. The overlap is readily evaluated as
\begin{eqnarray}
|\kappa|^2=|\langle \zeta|D(\alpha)| \zeta\rangle|^2 = \exp \left\{ - 
|\alpha|^2 \left[2N+1 +\sqrt{N(N+1)} \cos 2 \delta
\right]\right\} \label{ovsq}\;,
\end{eqnarray}
where $\delta=\arg(\zeta) -\arg(\alpha)$. 
By inserting the overlap in Eq. (\ref{KLmin}) we obtain the minimum detectable
$|\alpha|^2$. However, Eq. (\ref{ovsq}) shows a very strong dependence of
$|\alpha|^2_{\min}$ on the phase parameter $\delta$, which makes the
whole optimized scheme very unstable, namely one should know the phase
of perturbation very precisely in order to get a truly optimized
detection. Indeed, we have  
\begin{eqnarray}
|\alpha|^2_{\min} \stackrel{N\gg1}{\simeq} &
\Lambda(Q_0,\gamma^\star) / 4N  &\qquad {\rm for} \quad \delta=\pi/2 \\ 
|\alpha|^2_{\min} \stackrel{N\gg1}{\simeq}&
4 N \Lambda(Q_0,\gamma^\star) &\qquad {\rm for} \quad \delta=0  
\label{ovsq1}\;,
\end{eqnarray} 
with the second expression that shows an asymptotically divergent 
behavior in $N$.  \par
Let us now consider an entanglement-assisted scheme. We suppose you
have available a two-mode probe state $|\psi\rangle\!\rangle$ and
consider the configuration $U_\alpha=D(\alpha)\otimes{\mathbb I}$ in
which the displacement perturbs one mode, say $a$, and the other mode
is left unperturbed. As the probe state we consider the entangled
state from parametric downconversion of vacuum for finite gain---the
so-called ``twin-beam'' state
\begin{eqnarray}
|x\rangle\!\rangle= \sqrt{1-x^2} \: \sum_n \: x^n \: 
|nn\rangle\!\rangle \qquad 0\leq x < 1 \label{twb}\;,
\end{eqnarray}
where here $|nn\rangle\!\rangle=|n\rangle_a \otimes |n\rangle_b$. The twin-beam
in Eq. (\ref{twb}) has mean photon number  $N=2x^2/(1-x^2)$ and it is
achieved starting from the vacuum via the unitary evolution
$|x\rangle\!\rangle=\exp [x(a^\dag b^\dag - ab)]  
|0\rangle\!\rangle$. In order to evaluate the sensitivity, the main
task  is now to calculate  the overlap 
$|\kappa|^2=|\langle\!\langle x |U_\alpha|x\rangle\!\rangle|^2 $. We have 
\begin{eqnarray}
\kappa &=& (1-x^2) \: \sum_{m=0}^{\infty}\sum_{n=0}^{\infty}\: x^{m+n}
\: \langle\!\langle mm|D(\alpha)\otimes{\mathbb I}|nn\rangle\!\rangle=\nonumber \\ 
&=& (1-x^2)\: \sum_{n=0}^{\infty}\: x^{2n}\: \langle n|D(\alpha)
|n\rangle = (1-x^2)\: {\rm e}^{-\frac12|\alpha|^2}\: \sum_{n=0}^{\infty}x^{2n} 
L_n (|\alpha|^2)=\nonumber\\ &=& 
\exp\left[-\frac{|\alpha|^2}{2}\frac{1+x^2}{1-x^2}\right]=
\exp\left[-\frac{|\alpha|^2}{2}(N+1)\right]\:.
\label{kappad}
\end{eqnarray}
Equation (\ref{kappad}) implies for $|\alpha|^2_{min}$ the scaling
\begin{equation}
|\alpha|^2_{\min}\simeq\frac{ \Lambda (Q_0,\gamma^\star)}{N+1}\:,
\end{equation}
which is remarkably independent on the phase of perturbation, and thus
represents a robust bound to the sensitivity of a single-mode displacement.
\subsection{Perturbation made of a single-mode squeezing (phase-sensitive amplifier)}
The second kind of perturbation that we analyze  is the squeezing of a
single radiation mode, which is described by the squeezing operator
$S(\zeta)$ in Eq. (\ref{sqU}). Without loss of generality we
can consider $\zeta=\bar\zeta=r$ as real and use the notation $U_r$ to
indicate the transformation, namely $U_r=\exp \left[-i r
A\right]$, with $A=i/2(a^{+2}-a^2)$. The spectrum of $A$ is continuous
\cite{bollini} and extends from $-\infty$ to $\infty$: this means that
the eigenvalues of $U_r$ cover the whole unit circle. Therefore, it is
possible in principle to discriminate the perturbation exactly, using
as a probe either an eigenstate of the operator conjugated to $A$, or 
using a superposition of two or more eigenstates of $A$. However, 
analogously to the case of the
displacement, such probe states would be non normalizable and have
infinite energy, whence one must resort to physical approximations of
such states. For a coherent probe the overlap can be calculated through the overlap of 
the corresponding Wigner functions, giving as a result
\begin{eqnarray}
&&|\langle\alpha |U_r|\alpha\rangle|^2 =\nonumber\\ 
&&\qquad=\exp\left[ -\frac{2N\cos^2\phi(1-\cosh r-\sinh r)^2}{1
+\exp(2r)}-\frac{2N\sin^2\phi(1-\cosh r+\sinh r)^2}{1+\exp(-2r)}\right],
\end{eqnarray}
where $N=|\alpha|^2=\langle\alpha|a^{\dag}a|\alpha\rangle$
is the mean number of photons of the probe state. By expanding  
for small $r$ we have 
\begin{eqnarray}
|\langle\alpha |U_r|\alpha\rangle|^2 \simeq 1-Nr^2\:, \end{eqnarray}
and therefore the minimum detectable perturbation would be
\begin{equation}
r_{min}\simeq\sqrt{\frac{\Lambda(Q_0,\gamma^*)}{N}}.
\end{equation}
For a squeezed vacuum probe $S(\zeta)|0\rangle$ one has \cite{intj92}
\begin{eqnarray}
\kappa = \langle 0|S^{\dag}(\zeta)U_r S(\zeta)|0\rangle=
[\cosh r+2{\rm i}\sinh|\zeta|\cosh|\zeta|\sinh r\sin\psi]^{-\frac12}
\end{eqnarray}
where $\psi=\arg(\zeta)$ and correspondingly the minimum detectable r is given by:
\begin{eqnarray}
r_{min}=\cases{\log[\frac{1}{1-\Lambda(Q_0,\gamma^*)}\bigg[1-\sqrt{
\Lambda(Q_0,\gamma^*)\big(2-\Lambda(Q_0,\gamma^*)\big)}\bigg]]& 
\text{for $\sin\psi=0$,} \cr \sqrt{\frac{\Lambda(Q_0,\gamma^*)}{2}}
\frac{1}{N\sin\psi}& \text{otherwise,}}\label{rminsq1m}
\end{eqnarray}
with $N=\sinh^2|\zeta|$.
The bound in Eq. (\ref{rminsq1m}) strongly depends on the phase between the squeezing
perturbation and the squeezing of the probe, and therefore cannot be achieved
in practice without prior knowledge of the phase of the perturbation. \par
Let us now consider an entangled probe state in a twin beam state of
the form (\ref{twb}). The input-output overlap is calculated as follows
\begin{equation}
\kappa=\langle\!\langle x|U_r\otimes{\mathbb
I}|x\rangle\!\rangle=(1-x^2)\sum_{n=0}^{\infty}x^{2n} \langle 
n|U_r|n\rangle\:.\label{kappas}
\end{equation}
In order to calculate the matrix element $\langle n|U_r|n\rangle$ we
use the identities $S(r)={\rm e}^{\frac12\tanh(r) a^{\dag}}[\cosh(r)]^{-a^{\dag}a-\frac12}{
\rm e}^{-\frac12\tanh(r) a}$ and 
\begin{eqnarray}
{\rm e}^{-\frac12\tanh(r) a}|n\rangle =
&\sum_{l=0}^{[\frac{n}{2}]}\frac{[-\tanh(r)]^l}{2^l l!}
\sqrt{\frac{n!}{(n-2l)!}}|n-2l\rangle ,
\label{sq1n}
\end{eqnarray}
where $[m]$ indicates the integer part of $m$, 
and finally we get
\begin{equation}
\langle n|U_r|n\rangle=\frac{n!}{[{\rm
Ch}(r)]^{n+\frac12}}\sum_{l=0}^{[\frac{n}{2}]}\frac{(-1)^l [{\rm
sinh}(r)]^{2l}}{4^l (l!)^2 (n-2l)!}.
\label{eldiagsq}
\end{equation}
Using Eq. (\ref{eldiagsq}) we calculate $\kappa$ by means of Eq. (\ref{kappas})
\begin{eqnarray}
\kappa & = & (1-x^2)\sum_{n=0}^{\infty}x^{2n}\frac{n!}{[\cosh(r)]^{n+\frac12}}
\sum_{l=0}^{[\frac{n}{2}]}\frac{(-1)^l [\sinh(r)]^{2l}}{4^l (l!)^2 (n-2l)!}=
\nonumber\\ & = & \frac{(1-x^2)}{[\cosh(r)]^{\frac12}}\sum_{l=0}^{\infty}
\Bigg(-\frac{x^4{\rm sinh^2 (r)}}{4\cosh^2 (r)}\Bigg)^l
\frac{2l!}{l!^2}\sum_{n=0}^{\infty}\frac{(n+2l)!}{n!2l!}\Bigg(\frac{x^2}{\cosh(r)}
\Bigg)^n =
\nonumber\\  & &\qquad =
\frac{(1-x^2)}{[(x^4+1)\cosh(r)-2x^2]^{\frac12}}\nonumber.
\end{eqnarray}
Inserting this expression in Eq. (\ref{KLmin}) we have for $r_{min}$ the scaling law
\begin{equation}
r_{min} \simeq 2 \sqrt{\frac{\Lambda(Q_0,\gamma^\star)}{1-\Lambda(Q_0,
\gamma^\star)}}\frac{1}{\sqrt{N^2+2N+2}} \simeq \sqrt{\frac{ 
\Lambda(Q_0,\gamma^\star)}{1-\Lambda(Q_0,\gamma^\star)}}\frac{2}{N}\:. 
\label{rminsq}
\end{equation}
The same result is obtained by varying the phase of the squeezing
amplitude $\zeta$, thus confirming the robustness of the bound
(\ref{rminsq}) that is obtained using an entangled probe. 
\subsection{Perturbation made of a two-mode phase-shift}
The third problem we address is that of a perturbation induced by the
two-modes phase shift operator $a^{\dag}b+ab^{\dag}$, characterizing a mixer
(beam splitter) or a Mach-Zehnder interferometer.  This case differs from the 
previous ones in that the perturbation is represented by the two-modes 
unitary operator $V_\phi=\exp\{{\rm i}\phi(a^{\dag}b+ab^{\dag})\}$.
In this case the spectrum  is given by $\exp
\{{\rm i}m\phi\}$, with $m\in{\mathbb Z}$ (see e.g. \cite{job}). 
Therefore, if $\phi=(q /p)\pi$ with $q \in 2{\mathbb Z}+1$ and $p \in {\mathbb Z}$ 
(but this is a null-measure set of values of $\phi$)
then the optimal state is given by a superposition of two eigenstates 
of $V_{\phi}$ with eigenvalues differing by $\pi$. In the general case, the optimal 
state is any superposition of three or more eigenstates of $V_{\phi}$, such that the 
polygon of its eigenvalues on the unit circle encloses the origin 
\cite{entame}. Such optimal states are entangled, since they are obtained from 
the eigenstates $|n,d\rangle\!\rangle$ of $a^{\dag}a-b^{\dag}b$
\begin{equation}
(a^{\dag}a-b^{\dag}b) \: |n,d\rangle\!\rangle = d \:
|n,d\rangle\!\rangle,\quad
|n,d\rangle\!\rangle =\cases{|n+d\rangle |n\rangle & 
\text{for $d\geq 0$,}\cr |n\rangle |n+|d|\rangle & \text{for $d<0$,}}
\end{equation}
by the unitary transformation $\exp\{-(\pi/4)(a^{\dag}b-ab^{\dag})\}$.
Actually, the optimal states are far from being practically realizable.
However, we have proved that they are entangled, and this suggests to
explore the possibility of performing a reliable discrimination by
physically realizable entangled states.  For a twin-beam we have 
\begin{eqnarray}
\kappa&=&\langle\!\langle x|V_\phi|x\rangle\!\rangle=\nonumber\\
&=&(1-x^2)\langle\!\langle00|{\rm e}^{xab}{\rm e}^{{\rm i}\gamma_0
a^{\dag}b}{\rm e}^{\{\frac12\gamma_1 (a^{\dag}a-b^{\dag}b)\}}{\rm e}^{{\rm
i}\{\gamma_0 ab^{\dag}\}}{\rm e}^{xa^{\dag}b^{\dag}}|00\rangle\!\rangle,
\label{kappatmp}
\end{eqnarray} 
where $\gamma_0=\tan\phi$ and $\gamma_1=-\log(\cos^2\phi)$. After 
some algebra we get 
\begin{equation}
|\kappa|^2=\frac{1}{1+\frac{4x^2\sin^2\phi
}{(1-x^2)^2}}=\frac{1}{1+N(N+2)\sin^2\phi}\:.
\label{kappadmpf}
\end{equation}
The minimum detectable $\phi$, according to (\ref{kappadmpf}), is thus given
by
\begin{equation}
\phi_{min}=\arcsin
\left( \frac{\Lambda(Q_0,\gamma^\star)}{\sqrt{N(N+2)}}\right)
\simeq \frac{\Lambda(Q_0,\gamma^\star)}{N}\:.
\label{minphi}
\end{equation}
The scaling in Eq. (\ref{minphi}) does not depend on any parameter but
the energy of the input state. This should be compared with the sensitivity of 
the customary single-mode interferometry\cite{sha} based on squeezed
states, where the same scaling is achieved only for a very precise
tuning of the phase of the squeezing. This means that the
entanglement-assisted interferometry provides a much more reliable and
easily tunable scheme. 
\section{Implementation by difference-photocurrent interferometry}\label{s:diff}
\begin{figure}[h]
\psfig{file=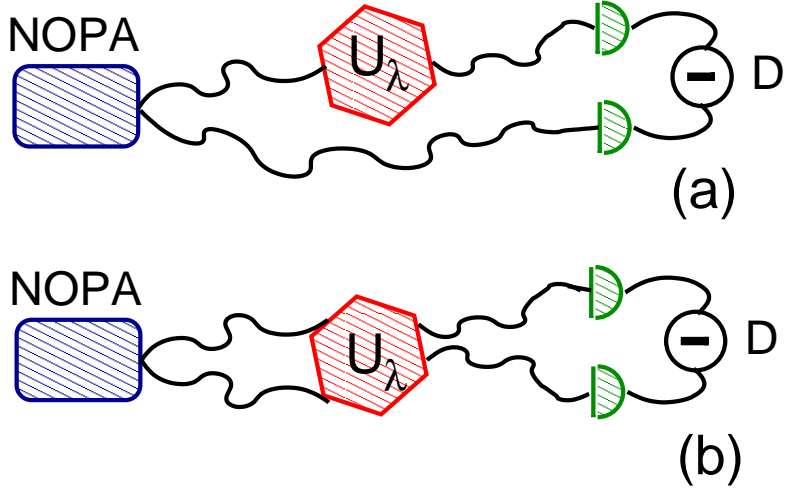,width=12cm}
\caption{Interferometric scheme to achieve ultimate bounds on precision by 
means of an entangled probe. The NOPA generates a twin-beam which 
may be subjected to the action of the unitary $U_\lambda$. At the output 
the beams are detected and the difference photocurrent is measured. For an 
unperturbed interferometer the output is again a twin-beam state, and
the scheme is designed in order to obtain a constant zero difference
photocurrent, whereas a perturbation $U_\lambda$ would produce fluctuations 
in the difference photocurrent. (a) general scheme for  single-mode 
perturbation; (b) scheme for two-mode
perturbation.}\label{f:interf}\end{figure}

In this section we present a concrete scheme for binary decision
based on an entangled probe. The scheme should be feasible with current
technology, and would allow to approach the ultimate precision bounds
that have been obtained in the previous sections. 
In Fig. \ref{f:interf} we show a schematic diagram of the
interferometric setup. The input state is the entangled twin-beam $|x\rangle\!\rangle$ 
produced by a nondegenerate optical parametric amplifier (NOPA).
Such entangled probe is possibly subjected to the action of the unitary
$U_\lambda$ (Figs. \ref{f:interf}a and \ref{f:interf}b describe the cases 
of a single mode and a two-mode perturbation respectively). At the output 
the two beams are detected and the difference
photocurrent $D=a^\dag a - b^\dag b$ is measured. If no perturbation occurs,
then the output state is still a twin-beam, and since $|x\rangle\!\rangle$ is an 
eigentstate of $D$ with zero eigenvalue we have a constant zero outcome for the 
difference photocurrent. On the other hand, when a perturbation occurs the output 
state is no longer an eigenstate of $D$, and we detect fluctuations
which signal the presence of the perturbation itself. The false-alarm
and the detection probabilities are given by 
\begin{eqnarray}
Q_0 &=& P (d \neq 0 | {\rm not} \: U_\lambda) \equiv 0 \\
Q_\lambda &=& P (d \neq 0 |  U_\lambda) = 1 - P(d\equiv 0 | U_\lambda)
\label{inQs}\;,
\end{eqnarray}
where the probability of observing zero counts at the output, after the action
of $U_\lambda$, is given by
\begin{eqnarray}
P(d\equiv 0 | U_\lambda) = \sum_n \: \left|\langle\!\langle n,n | U_\lambda |x\rangle\!\rangle\right|^2 
\label{xx} \label{pdl}\;,
\end{eqnarray}
since the eigenvalue $d=0$ is degenerate.
In this case the false-alarm probability is zero and therefore it is not necessary to
introduce an acceptance ratio. The scaling of the minimum detectable perturbation 
can be obtained directly in term of the detection probability $Q_\lambda$ by
Eqs. (\ref{inQs}) and (\ref{xx}). 
For the three transformations considered in the previous Section we have \\ 
\centerline{\bf Displacement}
\begin{eqnarray}
P(d=0 | \alpha\neq 0) = \exp \left(- |\alpha|^2(1+N) \right) I_0(|\alpha|^2 \sqrt{N(N+2)}) 
\quad\longrightarrow\quad |\alpha |^2_{min} \simeq \frac{\sqrt{Q_\lambda}}{N}
\label{intdis}\;
\end{eqnarray}
\centerline{\bf Squeezing}
\begin{eqnarray}
P(d=0 | r \neq 0) = 1 - r^2 N + O(r^2)
\quad\longrightarrow\quad r_{min} \simeq 
\frac{\sqrt{Q_\lambda}}{N}
\label{intsq}\;
\end{eqnarray}
\centerline{\bf Two-mode phase-shift}
\begin{eqnarray}
P(d=0 | \phi \neq 0) = 1 - \frac12 \phi^2 N^2 + O(\phi^2)
\quad\longrightarrow\quad \phi_{min} \simeq \frac{\sqrt{2 Q_\lambda}}{N}
\label{inttwm}\;.
\end{eqnarray}
On can see that in all  examples considered above, a realistic  interferometer 
based on a difference-photocurrent measurement  provides a precision
that re-scales with the energy in the same way as the ultimate bounds
obtained in the previous sections. 
\par
It is worth noticing that the experimental measurement of a modulated
absorption based on entanglement-assisted difference-photocurrent detection 
has been already performed using the entangled beam exiting an 
amplifier above threshold (parametric oscillator, OPO) \cite{exp}.
\section{Summary and conclusions}\label{s:outro} 
In this paper we have analyzed the effect of entanglement on the interferometric 
estimation of relevant quantum optical parameters such as displacing and squeezing 
amplitudes or interferometric phase-shift. We have evaluated the minimum detectable 
perturbation according to the Neyman-Pearson detection strategy, and have shown
that entanglement always improves the detection in comparison with conventional schemes.
In particular, for the case of estimation of the displacement and the squeezing amplitudes 
we have shown that the precision of the apparatus that use an
entangled probe is independent of the phase of the perturbation, and 
is therefore more robust and reliable of a simple scheme based on
single-mode probe states. Similarly, for the estimation of a two-mode
phase-shift, entanglement the interferometer  is much more stable when we use a
twin beam that when we use squeezed states. 
\par
Since the Neyman-Pearson detection strategy does not correspond to a realistic
detector, we proposed a feasible interferometric setup that is based on the
measurement of the difference photocurrent on an entangled
twin-beam. Remarkably, this scheme has the same energy-scaling of the
ultimate precision bound, and at the same time is very stable.
We conclude that the technology of entanglement can be of great help
in improving precision and stability of quantum interferometers.
\section*{Acknowledgments}
\par This work has been cosponsored by the Istituto Nazionale di
Fisica della Materia through the project PAIS-TWIN, and by 
Ministero dell'Universit\`a e della Ricerca Scientifica e
Tecnologica (MURST) under the co-sponsored project {\em Quantum
Information Transmission And Processing: Quantum Teleportation And
Error Correction}. G. M. D. acknowledges support by DARPA grant
F30602-01-2-0528.  

\end{document}